\documentclass[reprint,
 amsmath,amssymb,
 aps,pre,
]{revtex4-2}

\usepackage{subfigure}
\usepackage{makecell}
\usepackage{graphicx}% Include figure files
\usepackage{dcolumn}% Align table columns on decimal point
\usepackage{bm}% bold math
\begin{document}

\preprint{APS/123-QED}

\title{Percolation in a simple cubic lattice with distortion}% Force line breaks with \\

\author{Sayantan Mitra}
\author{Dipa Saha}
\author{Ankur Sensharma}
\email{itsankur@gmail.com}
 \affiliation{Department of Physics, University of Gour Banga, Malda - 732103, West Bengal, India.}%Lines break automatically or can be forced with \\

\date{\today}% It is always \today, today,

\begin{abstract}
Site percolation in a distorted simple cubic lattice is characterized numerically employing the Newman-Ziff algorithm. Distortion is administered in the lattice by systematically and randomly dislocating its sites from their regular positions. The amount of distortion is tunable by a parameter called the distortion parameter. In this model, two occupied neighboring sites are considered connected only if the distance between them is less than a predefined value called the connection threshold. It is observed that the percolation threshold always increases with distortion if the connection threshold is equal to or greater than the lattice constant of the regular lattice. On the other hand, if the connection threshold is less than the lattice constant, the percolation threshold first decreases, then increases steadily as distortion is increased. It is shown that the variation of the percolation threshold can be well explained by the change in the fraction of occupied bonds with distortion. The values of the relevant critical exponents of the transition strongly indicate that percolation in regular and distorted simple cubic lattices belong to the same universality class. It is also demonstrated that this model is intrinsically distinct from the site-bond percolation model.

\end{abstract}

\maketitle
\section{Introduction}
Percolation is a fundamental model of statistical physics introduced in 1957 \citep{Broadbent}. It is perhaps the simplest model to exhibit a non-trivial and rich critical behavior \cite{Isichenko}. Its appealing features have been continuously attracting researchers since its inception. Not surprisingly, therefore, the research in percolation has flourished in exploring its potential applications in many fields \cite{Ball, Dotsenko, Gruzberg, Derenyi, Coniglio, Anekal, King, Sapoval, Saberi3, Albano1, Takeuchi, Grassberger, Zhou}. At the same time, the model has given ample opportunities to theorists and mathematicians to address fundamental questions and resolve elusive challenges \cite{Saberi1}.

Two basic variants of this model are site percolation and bond percolation. In a classic site (bond) percolation problem, the sites (bonds) of a lattice can either be empty or occupied. Initially, a lattice with all empty sites (bonds) is considered, and the sites (bonds) are then occupied one by one with a probability $p$, called the occupation probability. If two neighboring sites (bonds) are occupied, they are said to be linked to each other. All these linked sites (bonds) form a cluster. For low $p$, many clusters of small size exist in the lattice. The clusters grow larger as $p$ increases and at a sufficiently high occupation probability, a giant cluster spans the lattice. For an infinitely large lattice, the spatial extent of the spanning cluster is also infinite. The first occurrence of such a cluster marks a phase transition and the corresponding occupation probability is called the percolation threshold $p_c$. There is also another model called site-bond percolation \cite{Hovi, Tarasevich}, in which both the sites and bonds are considered together and occupied independently to achieve spanning. Apart from these basic models, there exist numerous other models in literature such as directed percolation \cite{Broadbent,Takeuchi}, bootstrap percolation \cite{Adler}, explosive percolation \cite{Achlioptas,Riordan}, first passage percolation \cite{Hammersley}, and many more. The value of the percolation threshold depends on the type of the lattice (or network) as well as on the predefined rules of the process.

Physicists are generally more interested in characterizing phase transitions by determining relevant critical exponents. It is often observed that the different variants of percolation share almost the same values for the critical exponents despite having very different percolation thresholds \cite{Lorentz, Manna, Kundu1, Xiao}. These models are then said to belong to the same universality class. Although these results indicate that the values of the exponents depend primarily on the dimension of the lattice, there are instances of non-universality too in two dimensions \cite{Hassan, Kundu2}.

Natural systems are hardly perfectly ordered ones. Therefore, studying the percolation properties of regular lattices leaves a gap between ideal and real situations. To incorporate natural irregularities, a new percolation model in a distorted square lattice was proposed \cite{Sayantan}. In that model, the sites of a regular lattice are systematically but randomly dislocated from their original positions in a regular lattice. The nearest neighboring sites are connected only if their distance is less than a predefined value, called the connection threshold. The simulations were performed using the Hoshen-Kopelman algorithm \cite{Hoshen}. It was found that spanning becomes difficult with distortion and spanning is not possible even with $100\%$ sites occupied if the connection threshold is less than the lattice constant of the regular lattice.

In this work, we extend this model for a simple cubic lattice (SCL) with distortion. The simulations are performed with the Newman-Ziff algorithm \cite{Newman1, Newman2}, which is more powerful in characterizing the critical behavior of the percolation transition. It is observed that when the connection threshold is set equal to or greater than the lattice constant of the regular lattice, the percolation threshold of a distorted SCL increases with distortion. This behavior is similar to that of the distorted square lattice. However, when the connection threshold is less than the lattice constant, the percolation threshold first decreases and then increases with distortion. This is the most striking difference with the distorted square lattice, for which no spanning is possible if the connection threshold is less than the lattice constant. The similarity in the values of the critical exponents strongly suggests that the percolation in regular and distorted SCLs belong to the same universality class. We also demonstrate that percolation in distorted lattices can not be thought of as another manifestation of site-bond percolation; these two models are distinct.

\begin{figure*}
\subfigure[]{\includegraphics[scale=0.30]{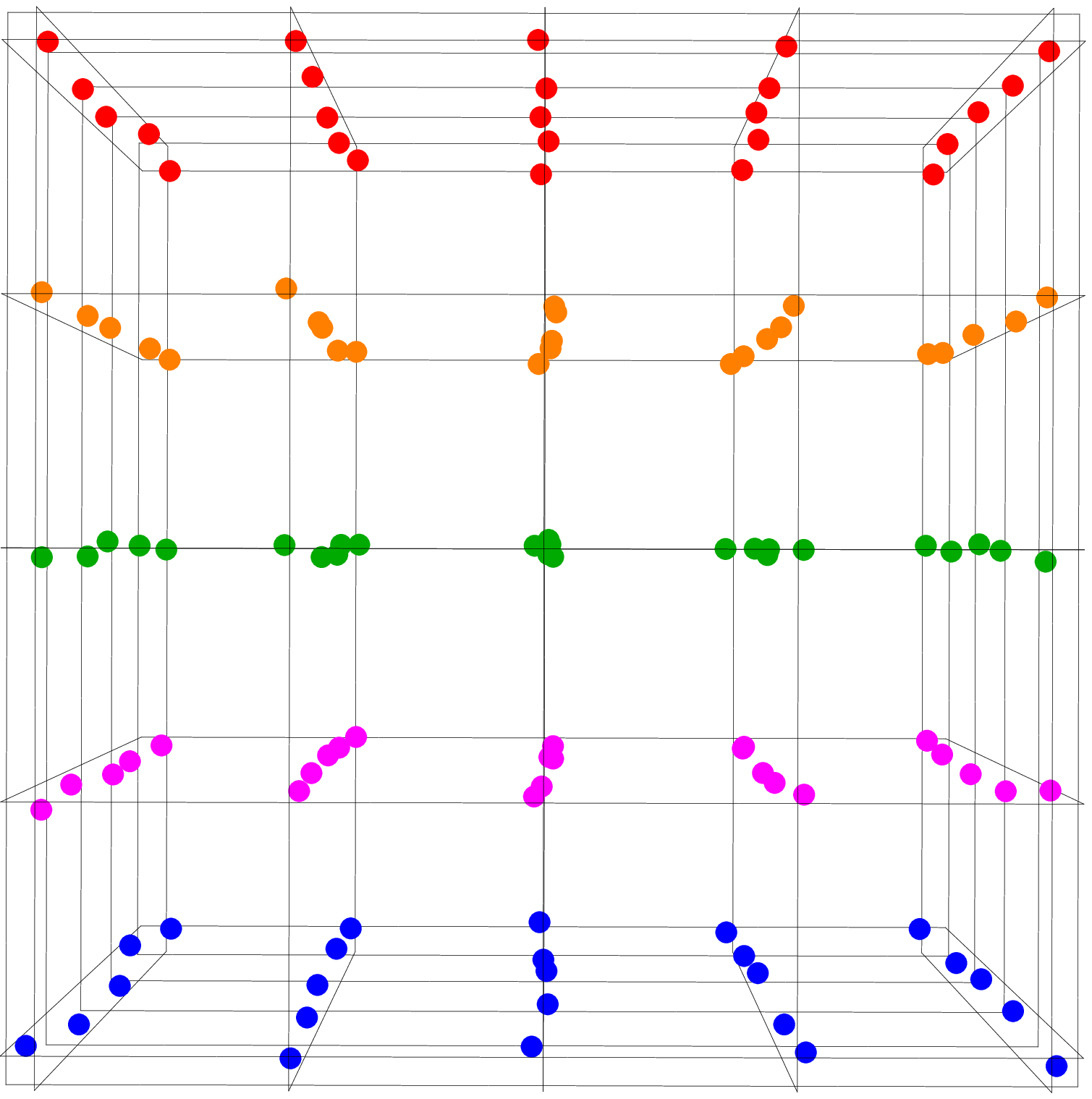}}\hspace{2cm}
\subfigure[]{\includegraphics[scale=0.60]{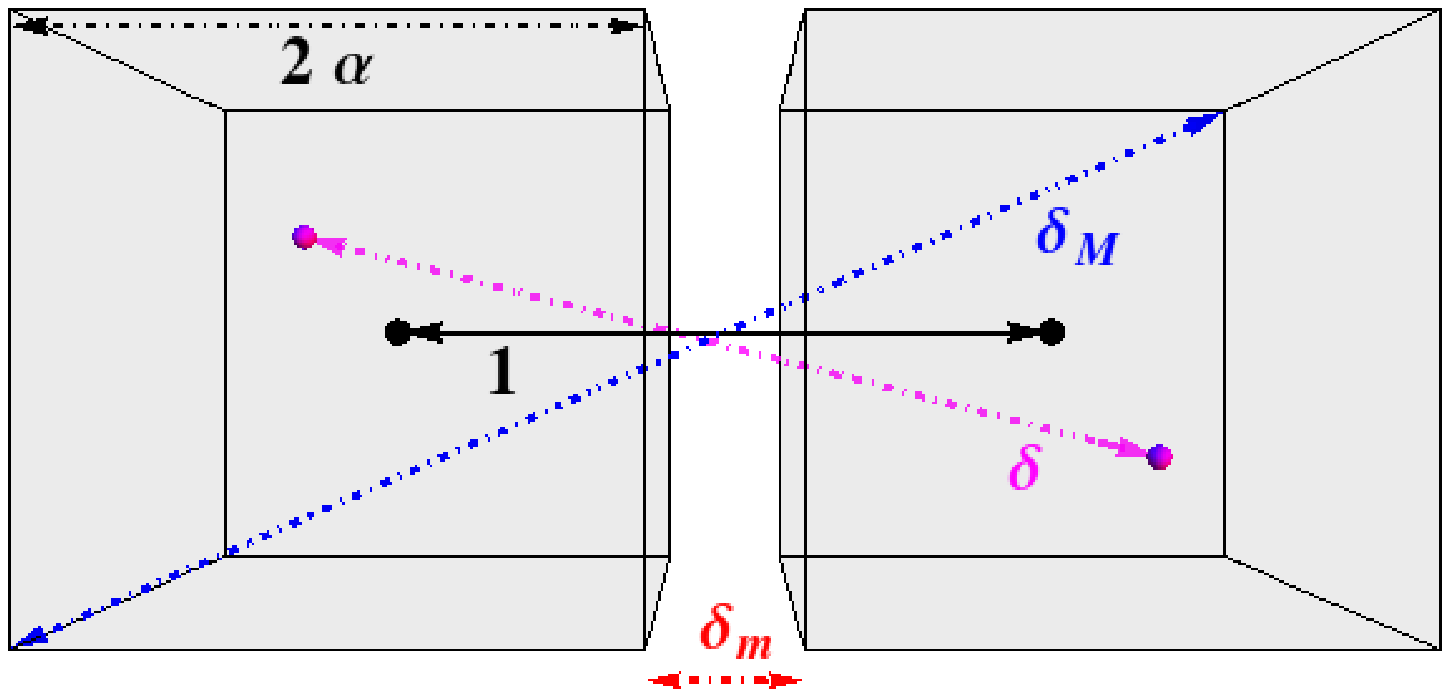}}
\caption{(Color online) (a) A realization of a $5\times 5\time 5$ distorted SCL. This lattice has been constructed from a regular lattice of unit lattice constant by dislocating the sites following $\alpha=0.05$. The viewing angle has been adjusted so that each row can be identified. (b) Magnified view of a pair of nearest neighbors in a distorted lattice. Each of them may be dislocated within a cube of length $2\alpha$ centered at the regular lattice positions. The distance between these two centers is $1$, while the distance between the sites is $\delta$, which lies in the range $\delta_m\le \delta \le \delta_M$.}
\label{fig:distlat}
\end{figure*}

The paper is organized as follows: In Sec.\ref{sec:lattice} and \ref{sec:cluster}, we describe the method of generating the distorted SCL and the process of cluster building, respectively, in our model. Sec. \ref{sec:pc_variation} illustrates the central result of this paper - the impact of distortion on the percolation threshold $p_c$. In Sec. \ref{sec:RE} we calculate $p_c$ of a finite lattice and show how it varies with the distortion parameter and the connection threshold. This is followed by the determination of an estimate of the percolation threshold of an infinite lattice $p_c^{\infty}$ for some combinations of these two parameters (Sec. \ref{sec:pc_inf}). Next, we characterize the percolation transition by determining the critical exponents in Sec. \ref{sec:exponents} and conclude that percolation in distorted and regular SCLs belong to the same universality class. Finally, in Sec. \ref{sec:distinction}, we demonstrate that the present model is distinct from the site-bond percolation model before summarizing our findings.
\section{The model}
\subsection{Generation of a distorted simple cubic lattice}\label{sec:lattice}
In this model, site-percolation is studied for a collection of sites arranged in a fashion that is nearly but not exactly an SCL. We call this a distorted simple cubic lattice. The lattice is distorted because the positions of the sites are not in general on the regular lattice points but are slightly dislocated. Although the amount and direction of these dislocations are random and independent for each lattice-point, control over the distortion has been enabled through the distortion parameter $\alpha$. The process of generating such a lattice is explained below.

To begin with, a regular SCL of sites with lattice constant $1$ is considered. A small cube of length $2\alpha$ is considered around each site keeping the site at the center of the cube. A given site is then dislocated to any position randomly within this small cube. This mechanism of shifting the positions of the sites is ensured by the following process. For each lattice point three separate random numbers, $r_x$, $r_y$, and $r_z$, each within the range $\{-\alpha,\alpha\}$, are generated for the shift of locations along $x,y,$ and $z$-directions respectively. Each site is then shifted accordingly so that the regular lattice position $(x,y,z)$ of a site changes to $(x+r_x,y+r_y,z+r_z)$. A distorted SCL is thereby realized. The amount of distortion can therefore be tuned by the parameter $\alpha$. 

Natural systems almost always have imperfections in their lattice structures. The idealized treatment of regular site percolation is therefore incomplete. The purpose of this study is to investigate the impact of these imperfections on percolation and the focus is therefore at low to moderate distortion ($\alpha\le 0.4$). Larger values of $\alpha$ would lead to almost a randomized array that falls outside the zone of interest of this work.

Fig. \ref{fig:distlat}(a) shows a schematic representation of a small distorted SCL. Note that, the  distance $\delta$ between a pair of nearest neighbor sites is not a constant. As shown in Fig. \ref{fig:distlat}(b), this distance may vary within the range $\delta_m\le \delta \le \delta_M$, where,
\begin{equation}\label{deltamin}
\delta_m=1-2\alpha,
\end{equation}
and
\begin{equation}\label{deltamax}
\delta_M=\sqrt{1+4\alpha+12\alpha^2}.
\end{equation}
The variation of the nearest neighbor distance $\delta$ is a key factor in the cluster building process which is explained next.

\subsection{Cluster building process}\label{sec:cluster}
In usual percolation, if two neighboring sites of a regular SCL are occupied, they are automatically directly linked and are always considered to be in the same cluster. In contrast, this is not guaranteed in a distorted SCL since the distances between the nearest-neighbor pairs are not the same anymore. The connection criterion for two occupied neighbors is set by introducing a connection threshold $d$. A direct link between two occupied neighboring sites exists only if they are close enough to ensure $\delta\le d$. Otherwise, the connection is broken even if both of them are occupied. Two limiting cases can readily be visualized from this criterion -- if $d<\delta_m$, no cluster formation is possible, and if $d\ge \delta_M$, the usual site-percolation scenario is restored. Therefore, the relevant ranges for $\delta$ and $d$ are the same.

It should be remembered that in the present model, the number of nearest neighbors of a site can not increase; it can only decrease. Like SCL, a site may be linked to at most $6$ other sites. In other words, a site can be directly linked to those sites which would have been its nearest neighbors in the regular lattice. The possibility of a direct link to other sites (by virtue of reduced distances due to distortion) is beyond the scope of the present study.

In this work, cluster numbering and identification have been done by the elegant Newman-Ziff (NZ) algorithm, which is known to give faster and more precise results as compared to other algorithms.
\section{Effect of distortion on the Percolation threshold}\label{sec:pc_variation}
One of the major goals of this study is to observe how the percolation threshold $p_c$ of an SCL is affected by distortion. It is anticipated that $p_c$ should depend on the distortion parameter $\alpha$ as well as the connection threshold $d$. To determine $p_c(\alpha,d)$, a distorted lattice with a given $\alpha$ needs to be generated, and a connection threshold $d$ needs to be set a priori. To demonstrate the variation of the percolation threshold with distortion, numerous combinations of $\alpha$ and $d$ must be taken into account, and $p_c(\alpha,d)$ must be calculated for each of them.

All the existing rigorous and detailed methods to calculate a precise value for the percolation threshold $p_c^{\infty}$ of an infinite lattice require a significant amount of computation time ( see, for example, \cite{Xiao}, \cite{Deng} and \cite{Wang}). It is therefore impractical to go through one of these procedures to determine $p_c^{\infty}$ for a large number of combinations of $\alpha$ and $d$. As a possible way out, an extremely simple and quick method of finding $p_c$ for a finite distorted SCL has been designed using the NZ algorithm. This enables us to show the variation of the percolation threshold in distorted SCLs with much less difficulty. Later, $p_c^{\infty}(\alpha,d)$ for some of the combinations of $\alpha$ and $d$ have been calculated. It has been revealed that these values are satisfactorily close enough to the corresponding estimates. We reiterate that the goal of this study is not to calculate very precise percolation thresholds up to several decimal places, but to understand the impact of $\alpha$ and $d$ on $p_c$.  This simple and quick estimation could therefore be very useful in gathering basic information of many other systems as well, before going for a detailed and rigorous method to obtain precise results for an infinite lattice.
\begin{figure*}
\subfigure[]{\includegraphics[scale=0.55]{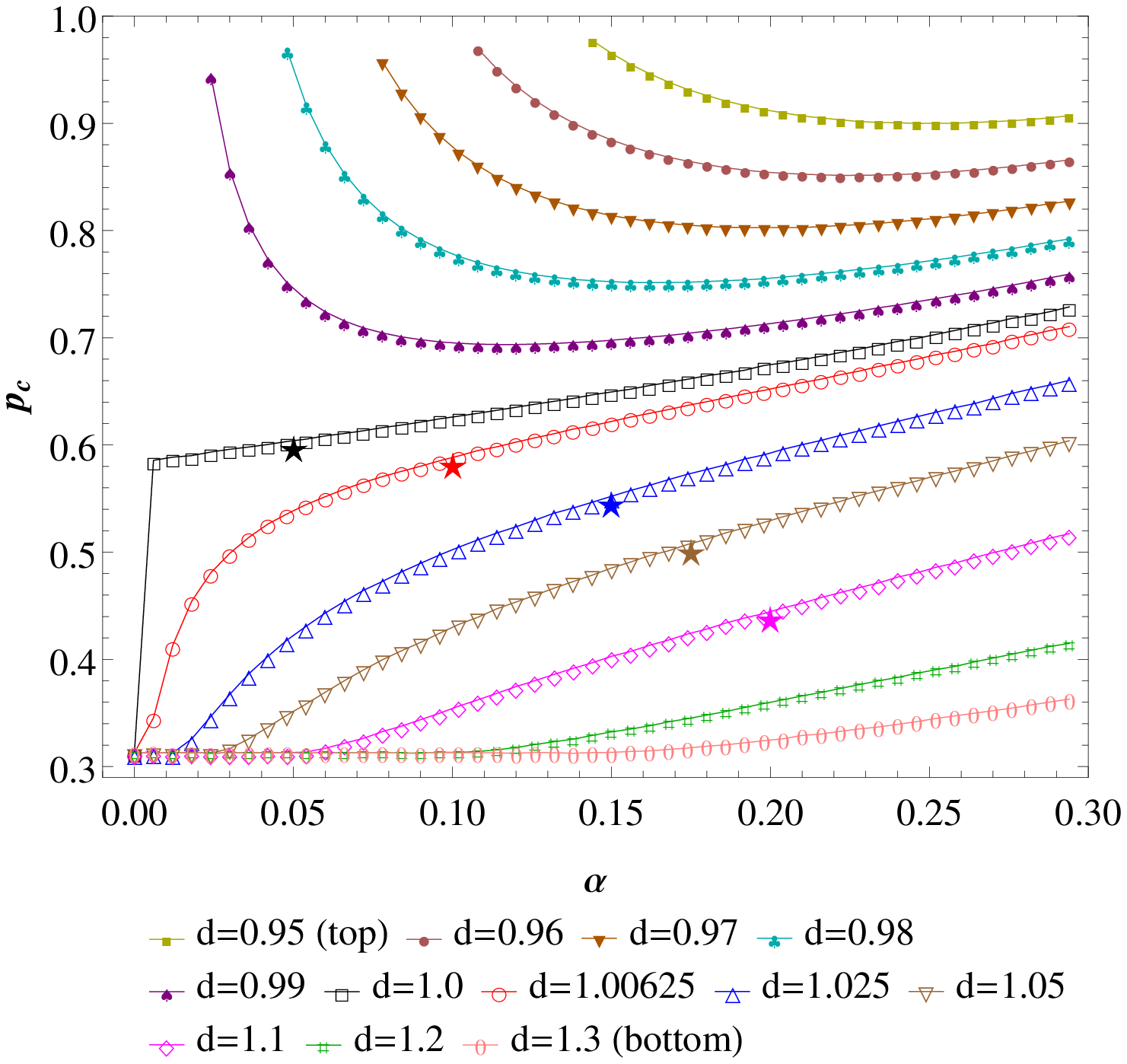}}
\subfigure[]{\includegraphics[scale=0.545]{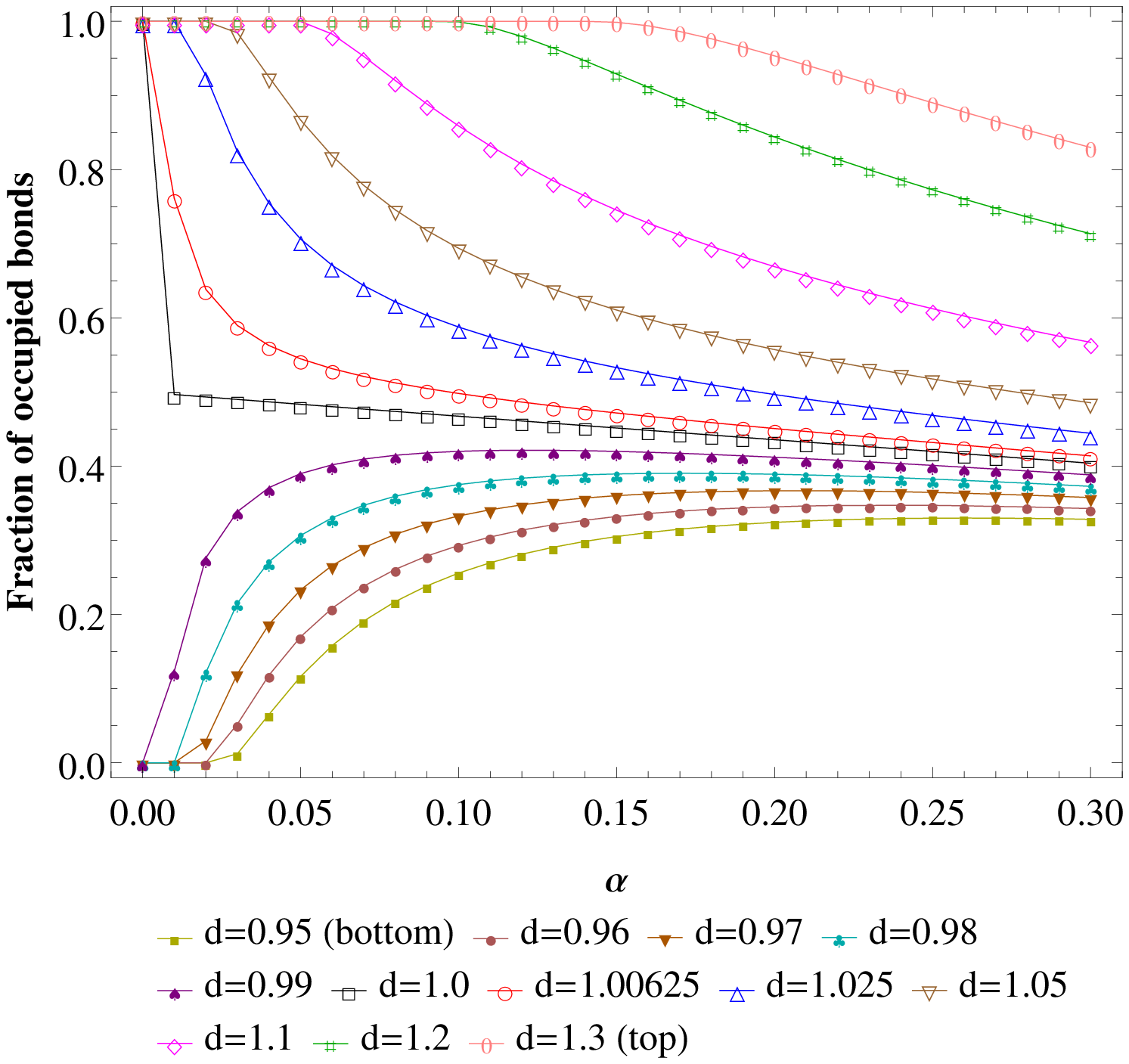}}
\caption{(Color online) (a) Variation of the percolation threshold $p_c$ for a finite lattice with distortion parameter $\alpha$. Each data point has been obtained by averaging over $10^3$ realizations of a distorted SCL of $L=2^7$. The curves are obtained simply by joining the points. Twelve curves are displayed for twelve different values of the connection threshold $d$. The five values of $p_c^{\infty}(\alpha,d)$ in Table \ref{tab:exponents} are indicated by stars. The natures of the curves are clearly different when $d\geq 1$ and $d<1$. (b) Plots of fraction of occupied bonds with $\alpha$ show exactly the opposite nature for the same set of values of $d$. Each data point has been obtained by averaging over $10^3$ realizations of a distorted SCL of $L=2^7$. Occupancy of the bonds depends only on the construction of the lattice and is independent of the occupancy of the sites.}
\label{fig:alphapc}
\end{figure*}
\subsection{Determination of $p_c(\alpha,d)$ for a finite lattice}\label{sec:RE}
First, a distorted SCL is generated with a fixed value of $\alpha$. As explained earlier, the distances $\delta$ between the neighboring sites are not fixed anymore. A connection threshold $d$ is then set to determine whether a given pair of occupied neighbors should be considered as directly connected (when $\delta\le d$)or not (when $\delta>d$). Starting with an empty distorted SCL, the following sequence of steps is executed. These operations follow the basic structure of the NZ algorithm.

\begin{itemize}
\item Each site is marked with a specific number and its position is recorded.
\item A randomized list is prepared to fix the order in which the sites are going to be occupied.
\item The sites are then occupied one by one as per the prepared list.
\item After occupying each site, the existence of links between occupied nearest neighbors are tested concerning the connection threshold $d$ previously set. The cluster structure and root-pointing are also examined and adjusted accordingly. (See ref. \cite{Newman2} for the details of the NZ algorithm.)
\item A checking is performed to detect the existence of a spanning cluster connecting two opposite sides of the lattice.
\item If a spanning cluster is not found, the next site of the list is occupied and the last two steps are repeated.
\item If a spanning cluster is found, no further sites are occupied and the occupation probability (the number of currently occupied sites divided by the total number of sites) is noted.
\end{itemize}

\begin{figure}
\includegraphics[scale=0.55]{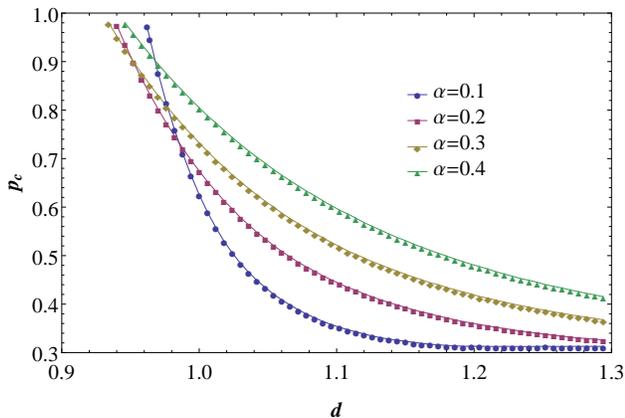}
\caption{(Color online) The (approximate) percolation threshold $p_c$ decreases with the connection threshold $d$. Each data point has been obtained by averaging over $10^3$ realizations of a distorted SCL of $L=2^7$. The curves are obtained simply by joining the points. Four curves are displayed for four different values of $\alpha$. Spanning becomes easier as $d$ increases since more links are allowed.}
\label{fig:dpc}
\end{figure}

The above scheme is repeated for several independent realizations of the distorted SCL with the same $\alpha$ and keeping the same $d$ for the connection criterion. The occupation probability $p$ for which the spanning cluster first appears is recorded for each of the realizations. After averaging over all the recorded values of $p$, an estimated value of $p_c$ is obtained for a fixed set of values of  $\alpha$ and $d$. Since this method involves considerably less computation time, we could estimate $p_c$ for nearly a thousand combinations of $\alpha$ and $d$. The calculations are done for distorted SCLs having $L=2^7$ sites along each side. This means that the occupation probability changes by an amount $\Delta p=1/L^3=4.77\times 10^{-7}$ when a new site is occupied. The results are illustrated in Fig. \ref{fig:alphapc}(a) and \ref{fig:dpc}. Every displayed point for $p_c$ has been obtained by averaging over $1000$ independent realizations of the distorted lattice with the same $\alpha$ and $d$. We have also checked that the variation of $p_c$ for smaller lattices such as $L=2^5,$ and $2^6$ are almost indistinguishably close to those for $L=2^7$. Therefore, as long as the lattice size is not too small, this estimate of $p_c$ is quite reliable.

Fig. \ref{fig:alphapc}(a) shows twelve curves, one each for a fixed value of $d$, with $\alpha$ varying in the range $\{0,0.3\}$, since we are interested in the low to moderate distortion regime. It is clear from the curves with $d>1$, that distortion causes difficulty in spanning, and as a consequence the percolation threshold is increased. Although this feature is somewhat similar to that of the distorted square lattices, some crucial facts and distinctions should be mentioned here.

A very reliable estimate of the percolation threshold for a regular SCL is known to be $p_{cu}=0.31160768(15)$ \cite{Xiao}. The curves with $d\ge 1$ start at this value (or, rather close to this value, as our calculation is for finite lattice) when $\alpha=0$. This is expected --  change in the connection threshold should not be manifested when distortion is absent. A glance at Eq. \ref{deltamax} reveals that $\delta_M=1$ for $\alpha=0$ and the condition $d\ge\delta_M$ to retain the regular percolation threshold is satisfied. This is also the reason why the curves for larger values of $d$ stays at $p_{cu}$ until $\alpha$ becomes large enough to ensure $d<\delta_M$.

A striking fact is noticed for $d=1.0$ ({\it i.e.} the connection threshold is equal to the lattice constant): even a slight distortion makes a huge impact on the percolation threshold. The value of $p_c$  stays close to $p_{cu}$ when $\alpha=0$, but it jumps to nearly twice of this value for a very small value of $\alpha$. After this initial jump, however, $p_c$ increases steadily with $\alpha$.

We wish to remark here that, no spanning cluster can be found for a distorted {\it square} lattice with $d\le 1.0$ even when all the sites occupied \cite{Sayantan}. Although there exist some pair of occupied neighboring sites for which $\delta<1$ (since $\delta_m=1-2\alpha$), the fraction of occupied bonds can never become large enough to span a distorted square lattice. In contrast, for a distorted SCL, we do obtain spanning for $d<1$ since a much lower fraction of occupied bonds (compared to a distorted square lattice) is required for spanning. In Fig. \ref{fig:alphapc}(a), five curves of $p_c(\alpha)$ are shown for $d=0.99,0.98,0.97,0.96,$ and $0.95$. These curves are of similar nature: an initial decline followed by a steady increment as $\alpha$ increases. This nature can be explained from Eq. \ref{deltamin} which says that the minimum distance $\delta_m$ between the nearest neighbors decreases with $\alpha$. For extremely low values of $\alpha$ ($\alpha<(1-d)/2$ in particular), $d<\delta_m$. Therefore, no bonds are occupied. As $\alpha$ becomes bigger than this value, some of the bonds start to be occupied. At a certain value of $\alpha$, the fraction of occupied bonds becomes sufficiently large enough to span the lattice, and we do obtain a finite value of $p_c$. When $\alpha$ is increased further, this fraction also increases (see Fig. \ref{fig:alphapc}(b)) which results in a decline in $p_c$. On the other hand, the average distance between the nearest neighbors slowly increases with $\alpha$, since $\delta_M$ increases with $\alpha$ faster than $\delta_m$ decreases. This reduces bond-occupancy. Therefore, the decreasing trend can not continue and $p_c(\alpha)$ starts to increase steadily after forming a minimum. Note that, no curves in Fig. \ref{fig:alphapc}(a) cross each other and the curves for $d<1$ always stay above the the curves for $d\ge 1$.

As explained earlier, two neighboring sites are directly linked, or, in other words, the bond between them is ``occupied" only if the distance between them is less than the connection threshold $d$. It should be noted that the occupancy of the bonds is fixed by the structure of the distorted lattice and is independent of whether the sites are occupied or not. Once a configuration is generated with fixed values of $\alpha$ and $d$, the number of occupied bonds gets fixed automatically. Thus, a spanning path is formed through collaboration between the occupied sites and the occupied bonds. Consequently, when more bonds are occupied, the site percolation threshold $p_c$ reduces. Therefore, the variation of the percolation threshold with distortion has a direct correspondence with the variation of the fraction of occupied bonds. Fig. \ref{fig:alphapc}(b) shows this dependence for the same set of values of $d$. It is clear that the nature of the curves for the corresponding values of $d$ is exactly reversed. For $d\ge 1$, $p_c$ always increases with $\alpha$, while the fraction of occupied bonds always decreases. The jump for $d=1$ is also present. When $d<1$, $p_c$ first decreases, forms a minimum, then increases steadily. Correspondingly, the fraction of bonds increases, forms a maximum, and then decreases steadily.

\begin{figure*}
\subfigure[]{\includegraphics[scale=0.80]{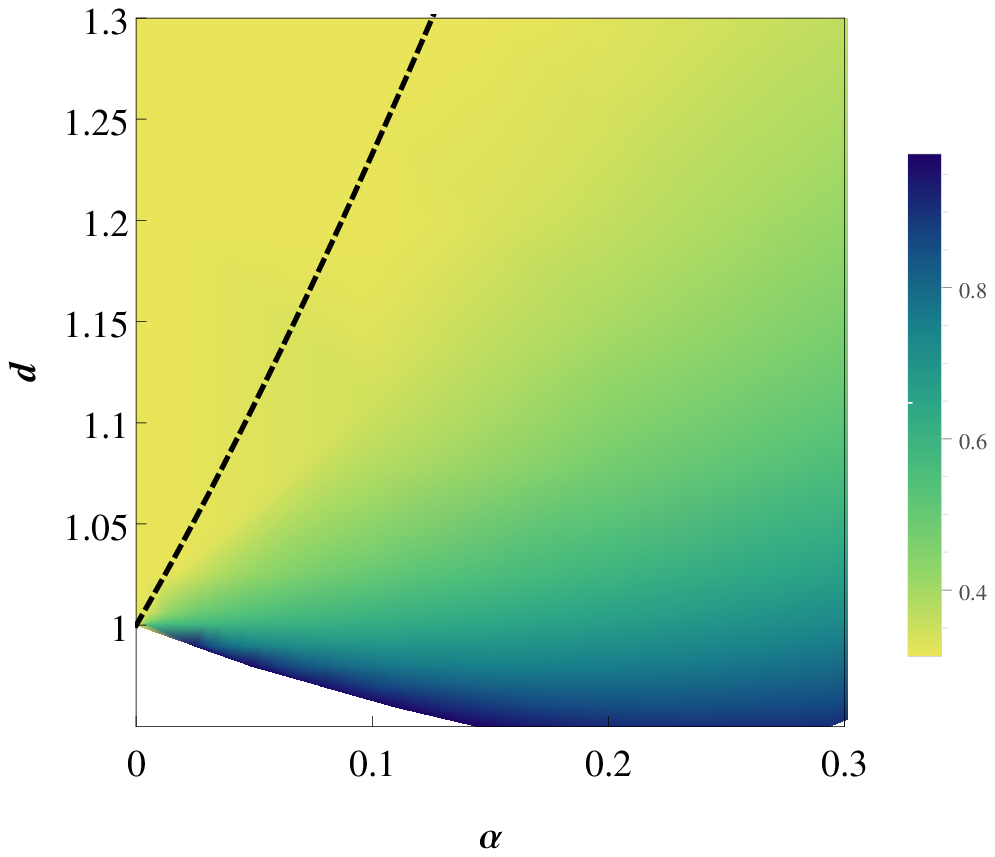}}
\subfigure[]{\includegraphics[scale=0.775]{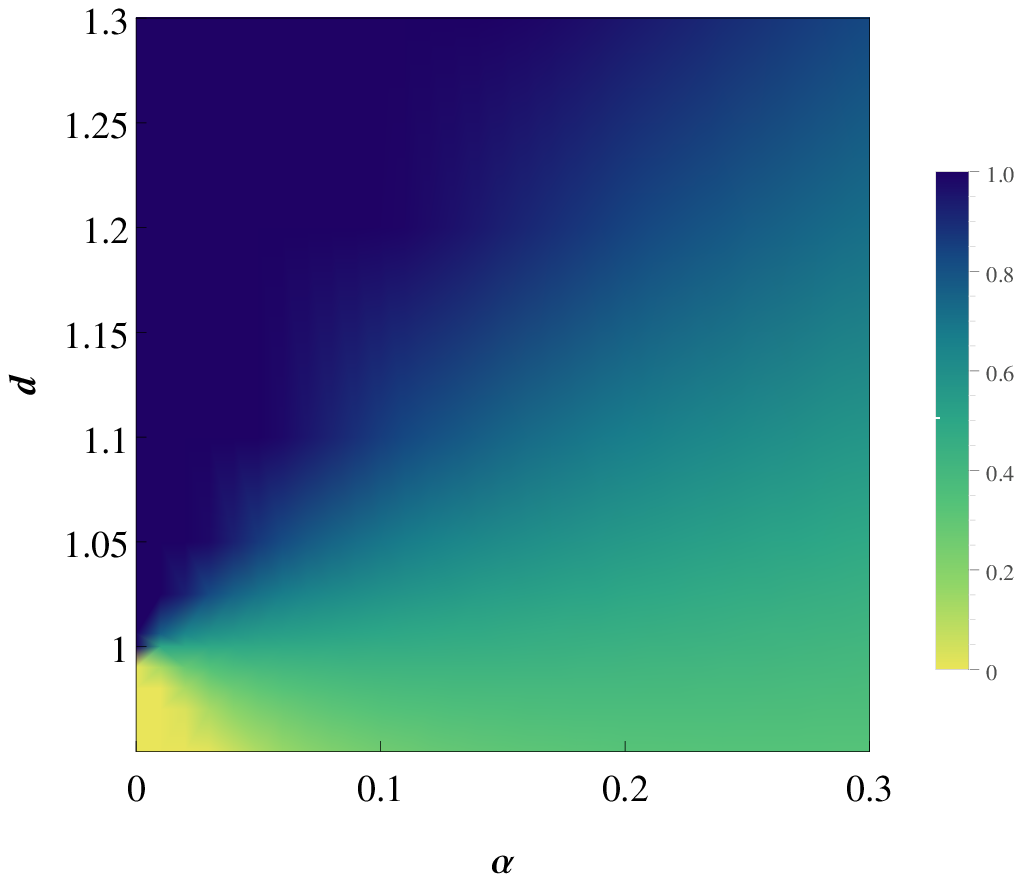}}
\caption{(Color online) (a) Variation of $p_c$ with $\alpha$ and $d$. The magnitude of $p_c$ is illustrated by color variation. The black dashed curve shows $\delta_M(\alpha)$ (Eq. \ref{deltamax}). In the region on the left of this curve, $d>\delta_M$. Consequently, $p_c$ remains constant at the value $p_{cu}$ in this region. The colorless portion at the bottom-left corner indicates that no spanning cluster can be found in this range. (b) Variation of fraction of occupied bonds with $\alpha$ and $d$. The low-valued yellow patch at the bottom-left corner explains the absence of spanning cluster in this regime.}
\label{fig:denplot}
\end{figure*}
Variation of $p_c$ with the connection threshold $d$ for different fixed values of $\alpha$ is shown in Fig. \ref{fig:dpc} Increase in $d$ means more connections are allowed and consequently existence of spanning cluster is more likely. It is therefore not surprising that $p_c$ reduces with $d$. Here also $p_c$ falls back to $p_{cu}$ for low $\alpha$ and high $d$ range when $d>\delta_M$ is satisfied (the curve for $\alpha=0.1$). There exist some crossings between the curves in $d<1$ range. This can be anticipated by carefully observing the curves of Fig. \ref{fig:alphapc}(a). For example, it is clear that, $p_c(\alpha=0.1,d=0.97)>p_c(\alpha=0.2,d=0.97)$, while $p_c(\alpha=0.1,d=1.1)<p_c(\alpha=0.2,d=1.1)$. Therefore, a crossing between the $p_c(d)$ curves for $\alpha=0.1$ and $\alpha=0.2$ is inevitable.

Fig. \ref{fig:denplot}(a) shows the dependence of the percolation threshold on both $\alpha$ and $d$. The magnitude of $p_c$ is represented by the color of the region - a darker shed means a higher value. As expected, $p_c$ is high in the lower part of the figure where $d$ is low. At the other extreme, $p_c$ remains close to the value $p_{cu}$. The black dashed curve, showing $\delta_M$ as a function of $\alpha$ (see Eq. \ref{deltamax}), marks the boundary of this region. On the left of this curve $d>\delta_M$, so every connection between the nearest neighbors is allowed. When this happens, the percolation thresholds of the distorted and regular SCLs must be the same. The colorless portion at the bottom-left corner reveals the fact that no spanning cluster can be found in this range of values of $\alpha$ and $d$. The variation of the fraction of the occupied bonds is shown in Fig. \ref{fig:denplot}(b). As expected, colors are reversed since an increase in the number of bonds results in a lower percolation threshold. The yellow patch at the bottom-left corner means that the fraction of occupied bonds is small in this region. This explains why no spanning cluster is found for very small values of $\alpha$ when $d<1$.

\subsection{Estimation of $p_c^{\infty}(\alpha,d)$}\label{sec:pc_inf}
The percolation threshold of an infinite lattice $p_c^{\infty}$ marks the occupation probability $p$ at which there is a sudden occurrence of an infinite cluster spanning the lattice. We calculate $p_c^{\infty}$ through the information of the spanning probability $S(p)$ -- the probability of occurrence of a spanning cluster for a given occupation probability $p$. For finite lattices, the monotonically increasing curves of $S(p)$ become steeper as lattice size is increased, and finally, for an infinite lattice, the curve approaches the shape of a step-function jumping from $0$ to $1$ at $p=p_c^{\infty}$. Therefore, these curves of spanning probability for different lattice sizes should intersect each other at $p=p_c^{\infty}$. 

\begin{figure}
\includegraphics[scale=0.55]{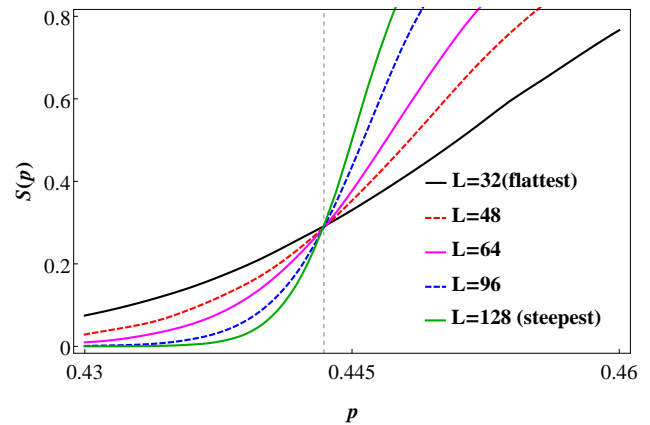}
\caption{(Color online) Interpolated plots of spanning probability for five different sizes of distorted SCL with $\alpha=0.2$ and $d=1.1$. The obtained $p_c=0.44342$ from all the intersection points is indicated by the vertical dashed line.}
\label{fig:spanprob}
\end{figure}

To find $p_c^{\infty}$ exploring the above concept, we generate $N=10^5$ independent realizations of a lattice of size $L$ with a fixed set of $\alpha$ and $d$. For each realization, some of the sites are occupied according to the occupation probability $p$. The number of realizations $n_s$ having spanning cluster (with free boundary conditions) is counted. The fraction $n_s/N$ approximately gives the spanning probability $S(p)$. In this way, the plots of $S(p)$ for lattice sizes $L=32,48,64,96,$ and $128$ have been generated. The obtained data points are interpolated to generate plots $S(p)$ and from the information of intersection points, $p_c^{\infty}$ has been estimated. Fig. \ref{fig:spanprob} shows the plots of $S(p)$ for $\alpha=0.2$ and $d=1.1$, the step-size of $p$ being $\Delta p=0.002$. Other combinations are not shown as they are similar in nature. It should be mentioned here that the accuracy is dependent on the sample size $N$, the number of lattice sizes considered, and the step-size $\Delta p$.

Using the above method, $p_c^{\infty}$ for a regular lattice has been found to be $0.311562(18)$, which is satisfactorily close to the currently accepted values \cite{Deng,Wang}. Table \ref{tab:exponents} shows $p_c^{\infty}(\alpha,d)$ for five combinations of $\alpha$ and $d$. These values are prominently indicated in Fig. \ref{fig:alphapc}(a). Note that, in Fig. \ref{fig:alphapc}(a) the variation of $p_c$ with $\alpha$ is shown and separate curves are obtained for different values of $d$. While plotting the points of Table \ref{tab:exponents}, only the values of $p_c^{\infty}$ and $\alpha$ have been provided. It is revealed that each $p_c^{\infty}(\alpha)$ is very close to the correct curve designated for $d$. This demonstrates the reliability of the quick estimation of the percolation threshold of a finite lattice described in Sec \ref{sec:RE}. 

\section{Critical exponents and universality class}\label{sec:exponents}
Although the mechanism of connectivity in distorted lattices is not the same as the ordinary percolation, and the percolation threshold is also different, it is expected that the values of the critical exponents would remain the same as long as the modified mechanisms are short-range. Nevertheless, it is worthwhile to verify this for the present model. Using the NZ algorithm, we focus on two important critical exponents $\beta$ and $\nu$ which are usually explored to decide on the universality class. Specifically, we determine the ratio $\beta / \nu$ and $\nu$ separately by two different methods. The obtained results indeed indicate that the values of these exponents do not depend on $\alpha$ and $d$. Also, these values are very close to those for ordinary percolation. Therefore, we conclude that percolation in distorted SCLs belongs to the same universality class as standard percolation. The same conclusion has been reached for percolation in distorted square lattices \cite{Sayantan, Jang}. It may therefore be stated that the distance-dependent spanning process of distorted lattices does not change the universality class.  

\begin{figure*}
\subfigure[]{\includegraphics[scale=0.56]{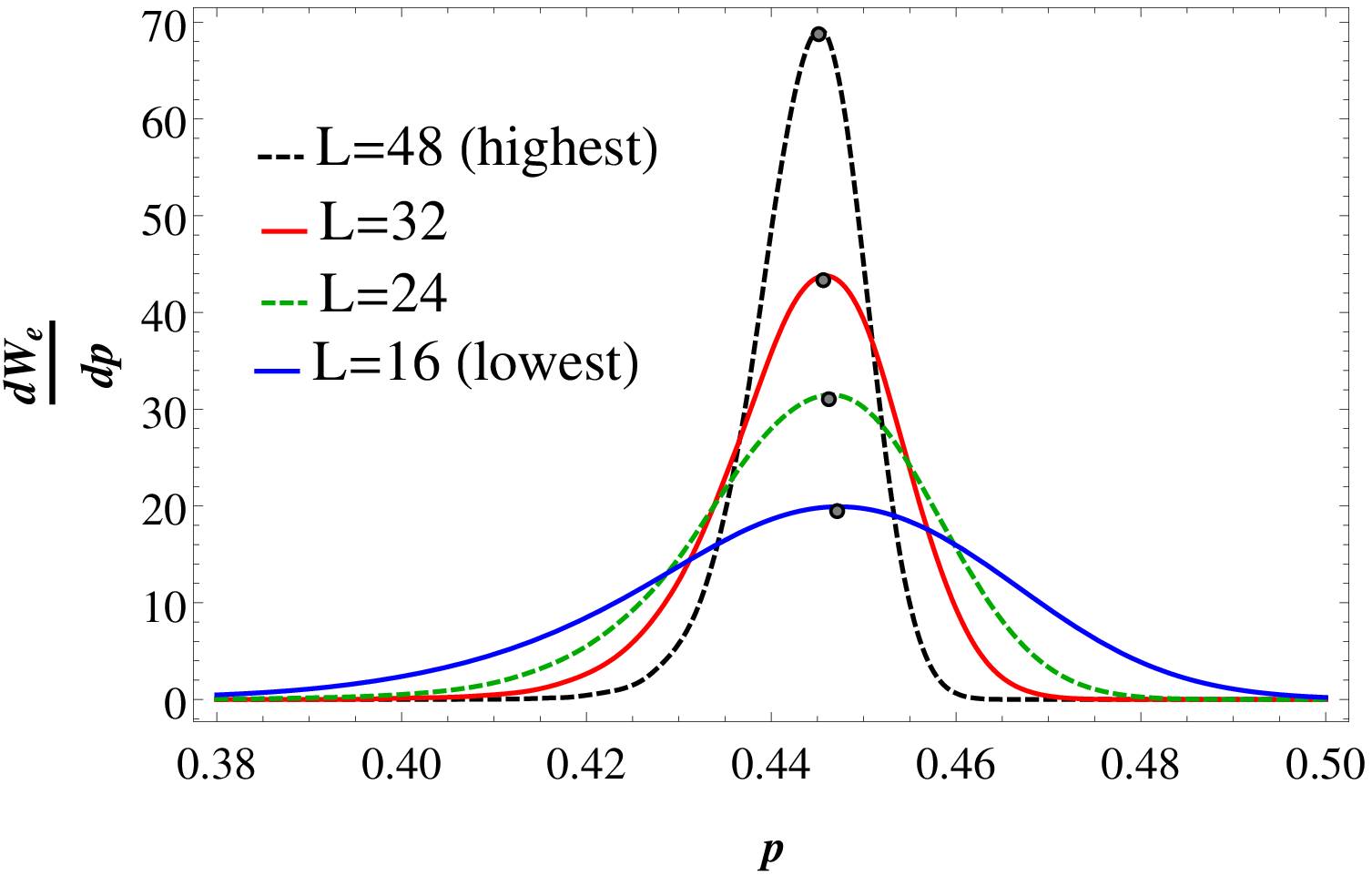}}
\subfigure[]{\includegraphics[scale=0.37]{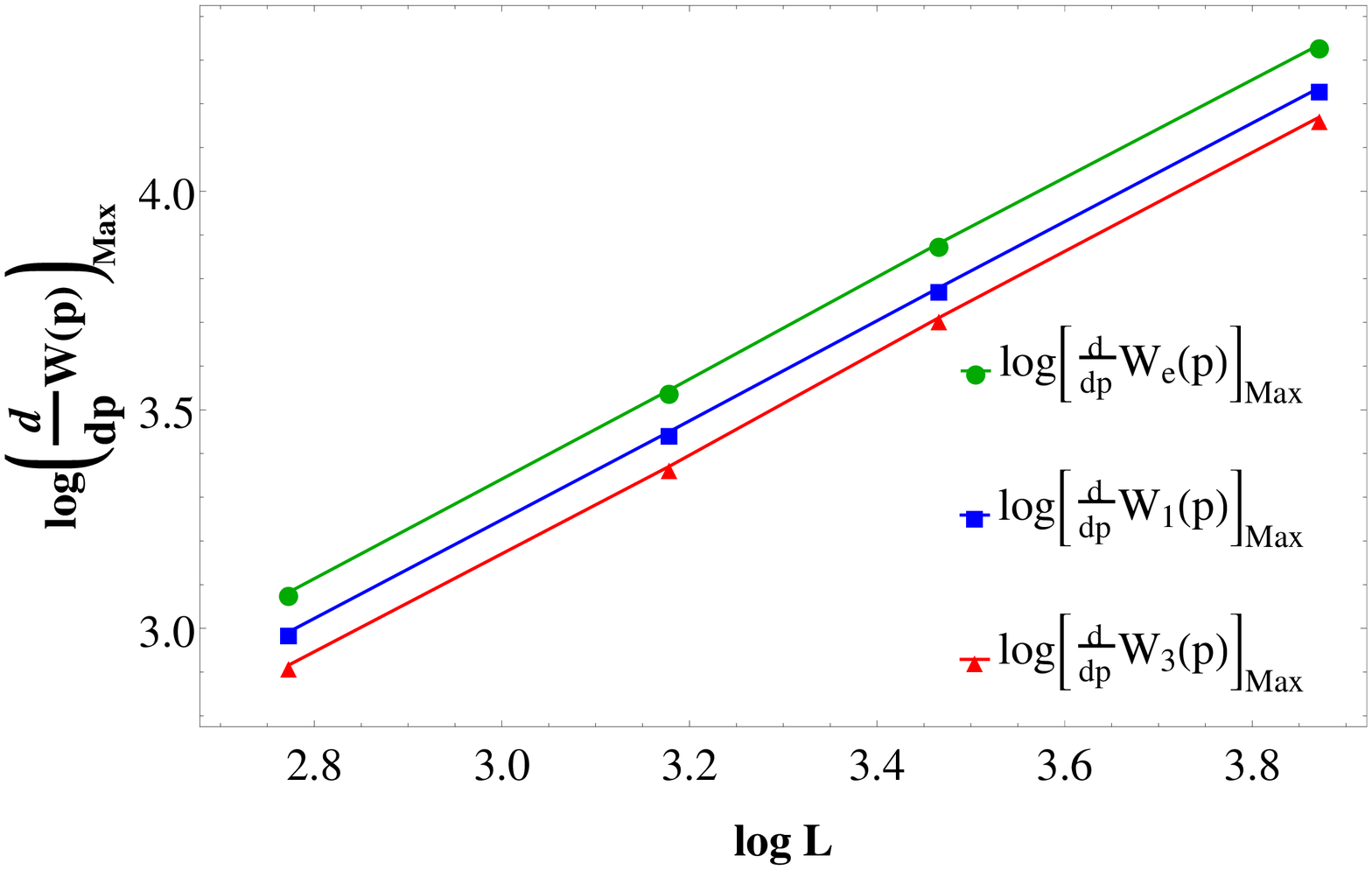}}\hspace{0.5cm}
\caption{(Color online) (a) The plots of $\Big(dW_e/dP\Big)$ for $L=16,24,32,48$. The maximum points used in the top most line of (b) are shown by dots.(b) The log-log plot of $\Big(dW/dP\Big)_{max}$ with $L$ for the three wrapping probabilities $W_1,W_e$, and $W_3$. For all the plots of this figure, $\alpha=0.2$ and $d=1.1$.}
\label{fig:1bynu}
\end{figure*}

\begin{table*}[ht]
\begin{tabular}{|c|c|c|c|c|c|c|c|}
\hline 
\rule[-1ex]{0pt}{2.5ex} $\alpha$ & $d$ & $p_c^{\infty}$ & $\beta /\nu$ & $d_f$ & \makecell{$\nu$ from \\ $[dW_3/dp]_{max}$} & \makecell{$\nu$ from \\ $[dW_1/dp]_{max}$} & \makecell{$\nu$ from \\ $[dW_e/dp]_{max}$}\\ 
\hline 
\rule[-1ex]{0pt}{2.5ex} 0.05 & 1.0 & 0.60254(3) & 0.468 & 2.532 & 0.889 & 0.883 & 0.883\\ 
\hline 
\rule[-1ex]{0pt}{2.5ex} 0.1 & 1.00625 & 0.58688(4) & 0.473 & 2.527 & 0.892 & 0.880 & 0.891\\ 
\hline 
\rule[-1ex]{0pt}{2.5ex} 0.15 & 1.025 & 0.55075(2) & 0.473 & 2.527 & 0.889 & 0.886 & 0.886\\ 
\hline 
\rule[-1ex]{0pt}{2.5ex} 0.175 & 1.05 & 0.50645(5) & 0.470 & 2.530 & 0.877 & 0.886 & 0.884\\ 
\hline 
\rule[-1ex]{0pt}{2.5ex} 0.2 & 1.1 & 0.44342(3) & 0.471 & 2.529 & 0.876 & 0.874 & 0.882\\ 
\hline 
\end{tabular} 
\caption{\label{tab:exponents}The values of the percolation threshold for infinite lattice and the critical exponents for five different combinations of $\alpha$ and $d$. Note that the critical exponents are essentially same while the percolation threshold varies significantly.}
\end{table*}

\subsection{Determination of $\nu$}\label{sec:nu}
The critical exponent $\nu$ can be estimated by evaluating the wrapping probability $W(p)$, which is defined as the probability of finding a cluster that wraps around the lattice. Depending on the direction of wrapping, a few variants of $W(p)$ are usually calculated \cite{Newman1, Newman2, Wang, Yang}. In this work, three such variants have been used for the estimation of $\nu$. At a given occupation probability $p$, these three variants are defined as (i) $W_1(p)$: the probability that a cluster wraps around the lattice along a specified axis, (ii) $W_e(p)$: the probability that a cluster wraps around the lattice along any one axis, and (iii) $W_3(p)$: the probability that a cluster wraps around the lattice along all three axes. The exponent $\nu$ can be found out from these wrapping probabilities through the scaling relation
\begin{equation}\label{nu}
\Big[\frac{d W(p)}{d p}\Big]_{max} \propto L^{1/\nu}
\end{equation} 

To find any of the above wrapping probabilities $W(p)$ at a given occupation probability $p$, one can use the convolution
\begin{equation}\label{conv}
W(p)=\sum\limits_{n=0}^N \Big(\begin{array}{c}
N \\ 
n
\end{array}\Big)    p^n(1-p)^{N-n}\langle W(n)\rangle,
\end{equation}
where, $W(n)$ is the wrapping probability when $n$ sites are occupied \cite{Newman2}. An advantage of this approach is that one can immediately find the first derivative without having to numerically differentiate. Differentiating Eq. \ref{conv},

\begin{equation}\label{dw}
\frac{d W(p)}{d p}= \sum\limits_{n=0}^N \Big(\begin{array}{c}
N \\ 
n
\end{array}\Big)    (n-Np)p^{n-1}(1-p)^{N-n-1}\langle W(n)\rangle
\end{equation}
We evaluate $W_1(n), W_3(n)$, and $W_e(n)$ numerically using the NZ algorithm with periodic boundary conditions. Hence using Eqs. \ref{conv} and \ref{dw} we calculate the derivatives. The plots of $dW_e/dp$ for system sizes $L=16,24,32$ and $48$ are shown for $\alpha=0.2$ and $d=1.1$ (Fig. \ref{fig:1bynu}(a)). Plots of other two wrapping probabilities look very similar. Locating $(dW/dp)_{max}$ and taking logarithm of Eq. \ref{nu}, $\nu$ can be evaluated from the gradient of the straight line (see Fig. \ref{fig:1bynu}(b)). Obtained values of $\nu$ from the three types of wrapping probabilities for five combinations of $\alpha$ and $d$ are summarized in Table \ref{tab:exponents}. These are close to the values found for a regular SCL \cite{Xiao, Wang, Borinsky}. The fluctuations are caused by the fact that $W_1(n), W_3(n)$, and $W_e(n)$ are evaluated only over $10^3$ configurations. However, even with this small sample size, the obtained values clearly suggest that $\nu$ has no dependence on $\alpha$ and $d$. 

\begin{figure*}
\subfigure[]{\includegraphics[scale=0.58]{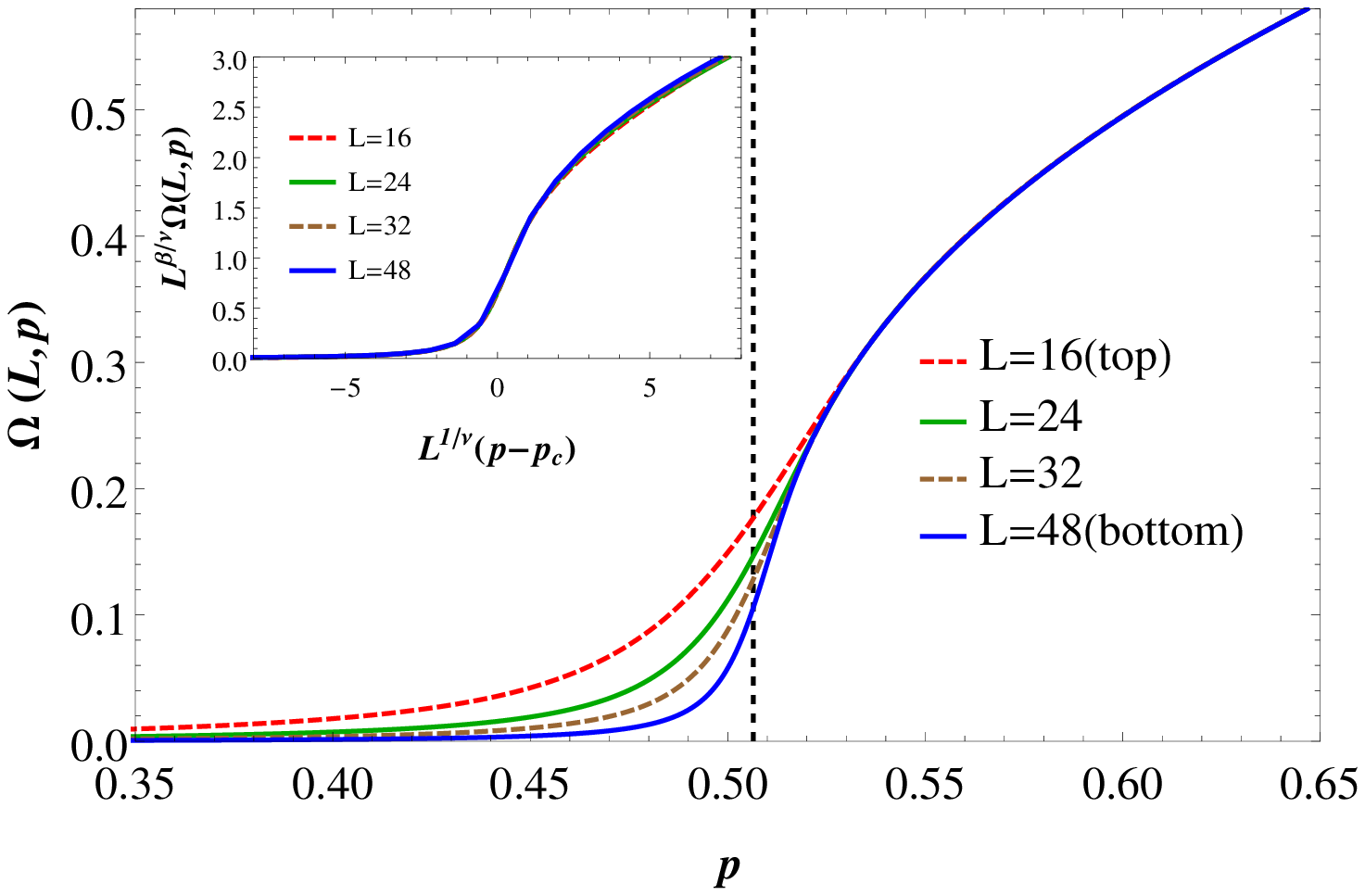}}
\subfigure[]{\includegraphics[scale=0.58]{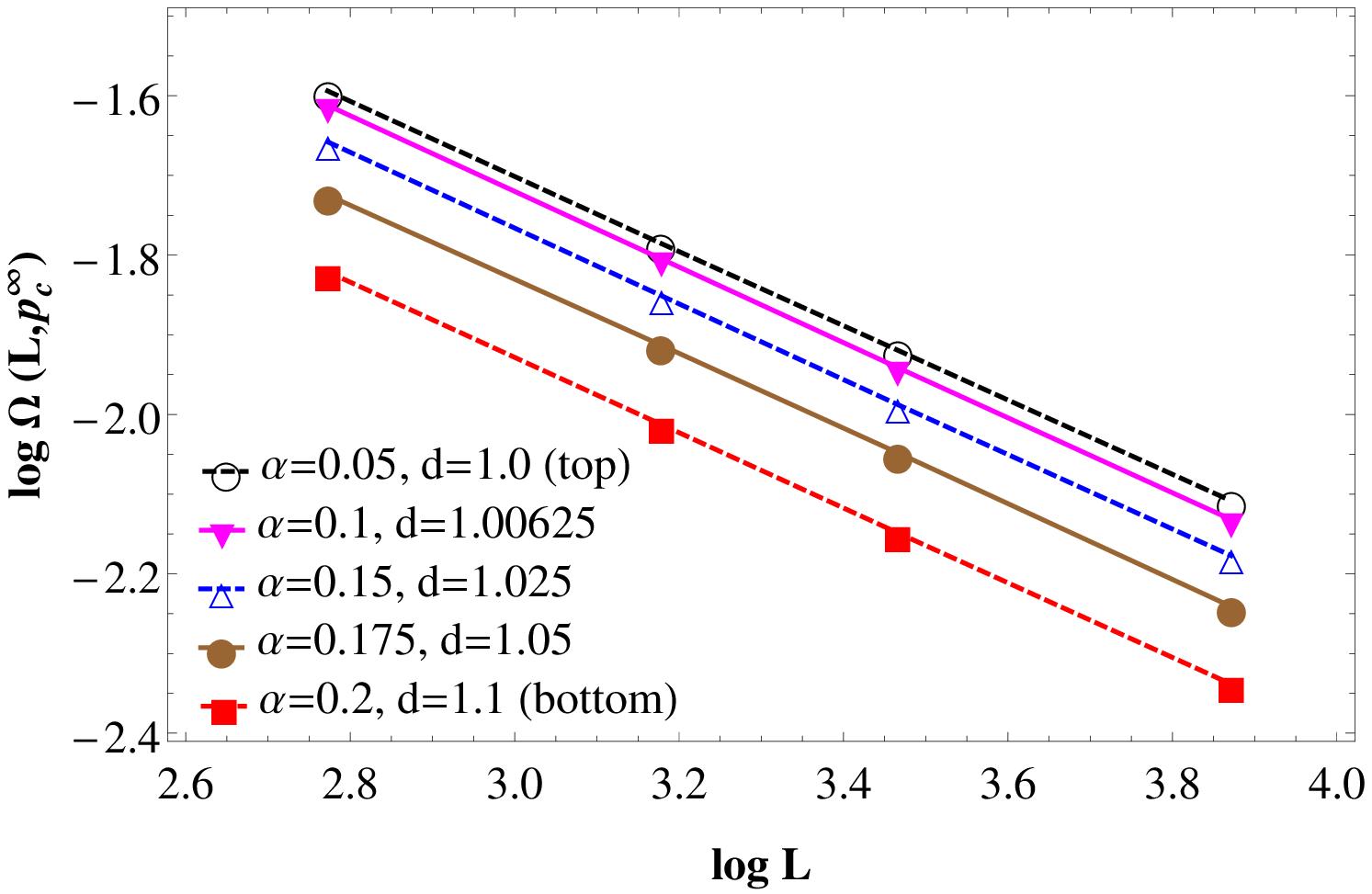}}
\caption{(Color online) (a) Plot of the order parameter $\Omega(L,p)$ for $L=16,24,32$, and $48$ with $\alpha=0.175$ and $d=1.05$. Corresponding $p_c^{\infty}$ is indicated by the dashed vertical line. Inset: Data collapse using $p_c=0.50645$, $\beta/\nu=0.470$, and $\nu=0.877$. (b) The log-log plot of the order parameter at percolation threshold $\Omega(L,p_c^{\infty})$ with lattice size $L$. The five straight lines correspond to the five sets of values of $\alpha$ and $d$ in Table \ref{tab:exponents}. The four points in each line correspond to $L=16,24,32$, and $48$. The lines are parallel since $\beta/\nu$ does not depend on $\alpha$ and $d$.}
\label{fig:betaneu}
\end{figure*}
\subsection{Determination of $\beta /\nu$}\label{sec:beta_by_nu}
The percolation order parameter is defines as
\begin{equation}\label{omega}
\Omega(L,p)=\frac{\langle S_{max}\rangle}{L^{d_m}},
\end{equation}
where, $S_{max}$ is the size of the largest cluster at occupation probability $p$, $L$ is the lattice size, and $d_m$ is the dimension of the space. Here $\langle\rangle$ denotes the configurational average. It is known that $\langle S_{max}\rangle$ is a fractal object at $p=p_c^{\infty}$ with fractal dimension $d_f$. Therefore, 
\begin{equation}\label{smax}
\langle S_{max}(p_c^{\infty})\rangle \propto L^{d_f}.
\end{equation}
The order parameter at the transition point is known to obey the scaling relation
\begin{equation}\label{betanu}
\Omega(L,p_c^{\infty})\propto L^{-\beta /\nu}.
\end{equation}
Using Eqs. \ref{omega}, \ref{smax}, and \ref{betanu} one can readily find
\begin{equation}\label{df}
d_f=d_m-\beta /\nu.
\end{equation}
With the information of  $p_c^{\infty}$ for five combinations of $\alpha$ and $d$, the ration $\beta /\nu$ is determined in the following manner. A distorted lattice of length $L$ with a given combination of $\alpha$ and $d$ is generated. To reduce the finite size effect, periodic boundary condition needs to be enabled. $S_{max}(p)$ is calculated for different occupation probabilities $p$. After averaging over $10^5$ such lattices, $\langle S_{max}\rangle$, and hence $\Omega(L,p)$ are determined. Variation of $\Omega(L,p)$ for $L=16,24,32$ and $48$ are shown in Fig. \ref{fig:betaneu}(a) for $\alpha=0.175$ and $d=1.05$. The curves are obtained by connecting closedly spaced data points. $p_c^{\infty}$ for this combination has already been determined to be $0.50645$, and shown by the dashed vertical line. Points of intersection of this line and the curves of $\Omega(L,p)$ give $\Omega(L,p_c^{\infty})$. The plot of $\log \Omega(L,p_c^{\infty})$ with $\log L$ should be a straight line with gradient $\beta /\nu$ (Eq. \ref{betanu}). In Fig. \ref{fig:betaneu}(b) five straight lines are shown for five $\{\alpha,d\}$ pairs of Table \ref{tab:exponents}. Each line is constructed by obtaining linear fit of four data points corresponding to $L=16,24,32,$ and $48$. $\beta /\nu$ can now be easily evaluated from the gradients of these lines. The lines of Fig. \ref{fig:betaneu}(b) are visibly parallel. It is not surprising therefore that the values of $\beta /\nu$ are very close to each other.  Table \ref{tab:exponents} shows these five values. The values of $d_f$ have been obtained from Eq. \ref{df} using $d_m=3$ for SCLs. These values are close to their corresponding values for a regular SCL \cite{Jan,Xiao,Deng,Wang}.  

It is known that the plots of $\Omega(L,p)$ of Fig. \ref{fig:betaneu}(a) should be collapsed when the horizontal and the vertical axes are scaled as $L^{1/\nu}(p-p_c)$ and $L^{\beta/\nu}\Omega(L,p)$, respectively. The exponent $\nu$ is calculated in the next section. Using $\beta/\nu =0.47$, and one of the obtained values $\nu=0.877$ for $\alpha=0.175$ and $d=1.05$ (see Table \ref{tab:exponents}), we indeed get a nice data collapse (See the inset of Fig. \ref{fig:betaneu}(a)).

The results of the critical exponents thus strongly indicate that the percolation in a regular and a distorted SCL belong to the same universality class.

\section{Distinction with site-bond percolation}\label{sec:distinction}
In site percolation, all the bonds are assumed to be preoccupied while a fraction of the sites is occupied as per occupation probability. Spanning occurs when a sufficient number of sites are occupied. The scenario is the opposite in the case of bond percolation, where all the sites are preoccupied and spanning is achieved by occupying bonds. There is another well-studied model called site-bond percolation, where neither all the sites nor all the bonds are preoccupied. Here, spanning is realized by randomly occupying a sufficient number of sites and bonds. Therefore, occupation probability for sites and bonds need to be specified separately.

The present model appears to be somewhat similar to the site-bond percolation model. The only difference is that the occupation of the bonds in distorted lattices is conditional, while for site-bond percolation it is random. In a distorted lattice, the number of links connecting the nearest neighbors is fixed by the two parameters $\alpha$ and $d$. But in site-bond percolation, the fraction of the occupied bonds is specified by the bond occupation probability $p_b$. A natural question therefore arises: are the site percolation thresholds of these two models the same when the same fraction of bonds are occupied (either randomly or conditionally)? If the answer is yes, one should conclude that the percolation in distorted lattices is just another manifestation of the site-bond percolation model. But we find that this is not the case.
\begin{figure}
\includegraphics[scale=0.55]{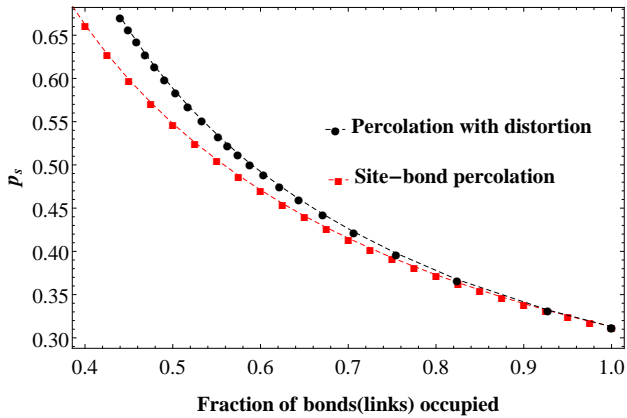}
\caption{(Color online) Demonstration of distinction between site-bond percolation in regular SCL and site percolation in distorted SCL. The site percolation thresholds for these two models are not the same when same fraction of bonds are occupied. Each data point reflects an average over $10^3$ realizations of a regular or distorted SCL with $L=128$.}
\label{fig:cppc}
\end{figure}
To establish this fact, we plot the site percolation thresholds for these two models with the fraction of the bonds (links) occupied (Fig. \ref{fig:cppc}). As expected, the results of these two models coincide when all the bonds are occupied. However, for less number of occupied bonds, the site percolation thresholds (in Fig. \ref{fig:cppc} it is written as $p_s$) are clearly different for these two models. It should be understood that, in a distorted lattice, the same fraction of links may be occupied for different combinations of $\alpha$ and $d$. We observe that $p_s$ is the same for all of them. But when the same fraction of bonds is occupied for site-bond percolation, $p_s$ is lower. So, if a fraction $f_0$ of bonds or links are occupied, one may write
\begin{equation}
\Big(p_s^{distorted}\Big)_{f_0}\ge \Big(p_s^{site-bond}\Big)_{f_0},
\end{equation} 
where the equality stands for $f_0=1$. To obtain the results of Fig. \ref{fig:cppc}, the quick estimation method described in Sec. \ref{sec:RE} has been used for both the models. The lattice size is $L=128$, and each point is obtained after averaging over $10^3$ configurations. The results of site-bond percolation are very close to those obtained by Gonz{\'{a}}lez {\it et. al.} \cite{Gonzalez} through a more rigorous method. This confirms that percolation in distorted lattices is distinct from the site-bond percolation model.

\section{Summary}
To summarize, we have studied site percolation in distorted SCLs. A distorted SCL is prepared from a regular SCL (lattice constant = $1$) by dislocating its sites randomly but systematically. The amount of distortion is tuned by the distortion parameter $\alpha$. The distances between the neighboring sites of the distorted lattice are no longer the same, and two occupied neighboring sites are directly linked to each other if their distance is less than the connection threshold $d$. We develop a method to find the percolation threshold $p_c$ for a finite lattice (plots are shown for $L=2^7$) and later confirm that this estimate is satisfactorily close to the actual percolation threshold $p_c^{\infty}$. The results are obtained by incorporating the distance-dependent connectivity of the neighboring sites into the Newman-Ziff algorithm. Our major findings are listed below.
\begin{itemize}
\item When the connection threshold $d$ is fixed at a value $\ge 1$, the presence of distortion makes spanning difficult as manifested by the increment in $p_c$ with $\alpha$. In particular, when $d$ is set to be equal to the lattice constant of the regular SCL, a discontinuous jump in $p_c$ is noted for a very small distortion.
\item However, for $d<1$, $p_c$ first decreases then increases steadily with $\alpha$. This is a prominent difference with the results of a distorted square lattice, where no spanning cluster can be found for $d<1$.
\item On the other hand, $p_c$ always decreases with $d$ for a fixed value of $\alpha$. Therefore, the percolation threshold in distorted SCL depends heavily on the interplay between the two parameters $\alpha$ and $d$.
\item To characterize the percolation transition, we calculate the critical exponents $\beta$, $\nu$, and $d_f$, and find that they are very close to their currently accepted values for regular lattices. Although a considerable modification is enforced in the percolation process, and there is a significant change in the percolation threshold, the critical exponents remain essentially the same. Therefore we conclude that percolation in regular and distorted SCLs belong to the same universality class.
\item Percolation in a distorted lattice appears to have some similarity with the site-bond percolation model. We have conclusively demonstrated that these two models are distinct from each other.

\end{itemize}
The goal of this study is to correctly characterize the percolation transition under the influence of distortion. Determination of the accurate percolation threshold and the critical exponents up to several decimal places is not aimed. We believe that the obtained results are sufficiently precise to reach the conclusions of this work.

The fact that distortion significantly impacts the percolation process, has the potential to generate considerable scientific interest in the future. To proceed further with this model, distorted versions of other Bravais lattices may be investigated. It would also be interesting to see the effects of relaxing the number of nearest neighbors of a site and identifying them only in terms of their distances from the site. This scenario will be prominent in highly distorted 2D and 3D lattices as well as in networks such as random geometric graphs. The works on percolation with extended neighborhood \cite{Xun1, Xun2, Xun3} may be useful in this regard.

\section{Acknowledgments}
The authors thank Akhileswar Prasad and Bishnu Bhowmik for illuminating discussions. The computation facility availed at the Department of Physics, University of Gour Banga is gratefully acknowledged.

\end{document}